\documentclass[journal=ancham,manuscript=article]{achemso}
\pdfoutput=1 
\usepackage[T1]{fontenc} 
\usepackage[normalem]{ulem}

\author{Julius Sefkow-Werner}
\altaffiliation{ Univ. Grenoble Alpes, CNRS, Grenoble INP, LMGP, 38000 Grenoble, France}
\author{Elisa Migliorini}
\email{elisa.migliorini@cea.fr}
\author{Catherine Picart}
\affiliation{ BRM ERL 5000 CEA/CNRS/UGA,France}
\author{Dwiria Wahyuni}
\altaffiliation{ Current address: Tanjungpura University, Pontianak, Indonesia}
\author{Ir\`ene Wang}
\author{Antoine Delon}
\affiliation[UGA]
{ LIPHY, Universit\'e Grenoble Alpes and CNRS, F-38000 Grenoble, France}
\email{antoine.delon@univ-grenoble-alpes.fr}

\title[photobleaching Fluctuation Fluorescence Spectroscopy]{Combining fluorescence fluctuations and photobleaching to quantify surface density}

\abbreviations{FCS,ICS,FFS}

\begin{document}

\begin{tocentry}

\includegraphics[width=1\textwidth]{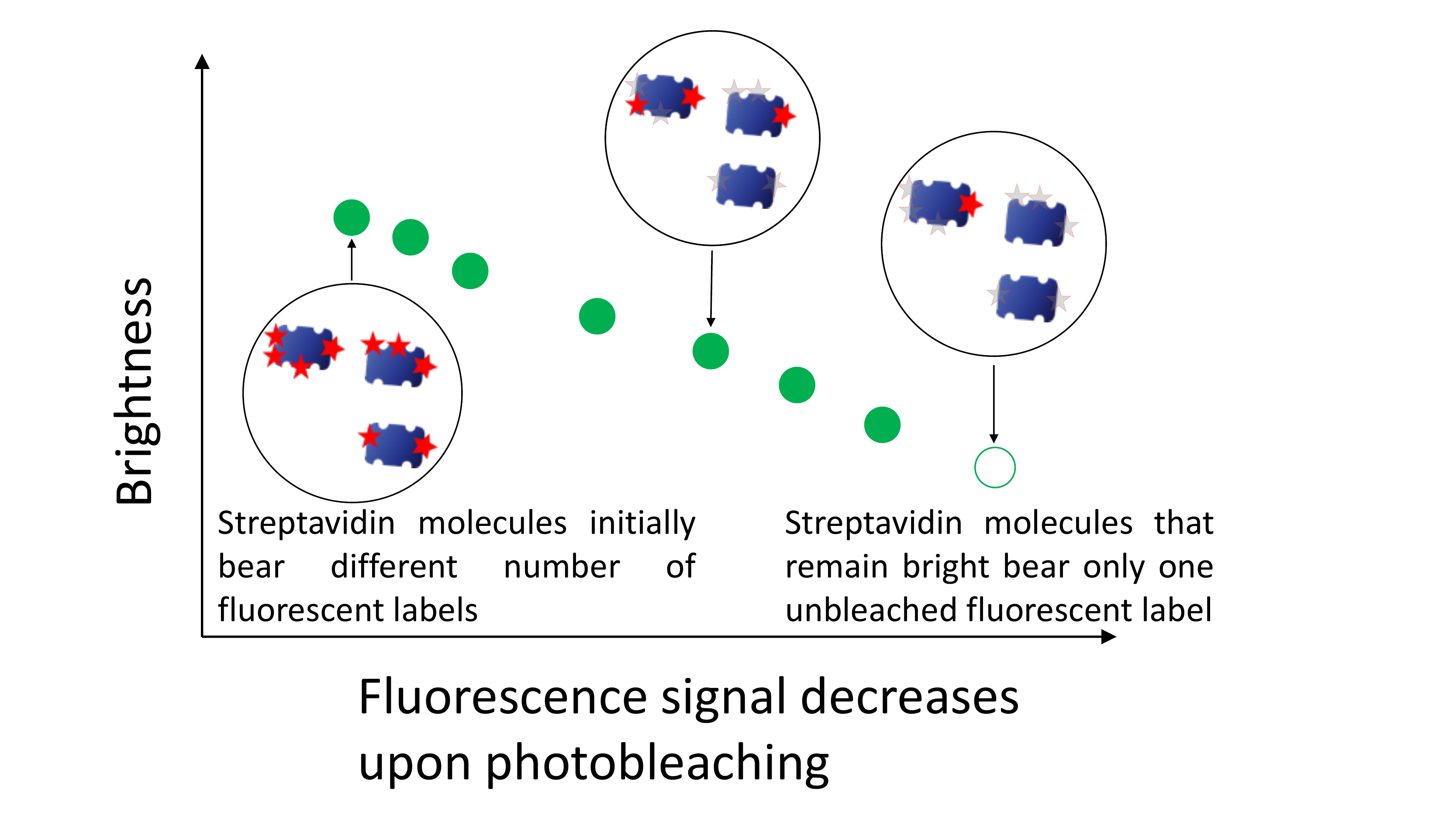}

\end{tocentry}

\begin{abstract}
We establish a self-calibrated method, called \emph{pbFFS} for \textit{photobleaching Fluctuation Fluorescence Spectroscopy}, which aims at characterizing molecules or particles labelled with an unknown distribution of fluorophores. Using photobleaching as a control parameter, pbFFS provides information on the distribution of fluorescent labels and a more reliable estimation of the absolute density or concentration of these molecules. 
We present a complete theoretical derivation of the pbFFS approach and experimentally apply it to measure the surface density of a monolayer of fluorescently tagged streptavidin molecules that can be used as a base platform for biomimetic systems. The surface density measured by pbFFS is consistent with the results of spectroscopic ellipsometry, a standard surface technique. However, pbFFS has two main advantages: it enables \textit{in situ} characterization (no dedicated substrates are required) and is applicable to low masses of adsorbed molecules, as demonstrated here by quantifying the density of biotin-Atto molecules that bind to the streptavidin layer. Besides molecules immobilized on surfaces, we also applied pbFFS to molecules diffusing in solution, to confirm the distribution of fluorescent labels found on surfaces. Hence, pbFFS provides a set of tools to investigate molecules labelled with a variable number of fluorophores, with the aim to quantify either the number of molecules or the distribution of fluorescent labels, the latter case being especially relevant for oligomerization studies.
\end{abstract}

\section{Introduction}
Biomimetic approaches are popular in medical applications and cellular studies. As the extracellular matrix (ECM) plays a complex role on cell response to drugs, growth factors and morphological cues \cite{Martino2015}, it is advantageous to design biomaterials mimicking the natural environment of cells in the body for enhancing the efficiency of biomedical products. Moreover these biomaterials can bring a deeper understanding of the influence of selected components of the ECM on cellular processes such as proliferation, migration and differentiation \cite{Migliorini2020}. Developing these platforms requires a precise control of the immobilized compounds. Standard surface techniques include spectroscopic ellipsometry and quartz crystal microbalance with dissipation monitoring (QCM-D)\cite{SefkowWerner2020, Migliorini2018}. However, these techniques are \textit{ex situ} since the platforms have to be built on auxiliary substrates, which may affect the functionalization process. Moreover, low masses of adsorbed molecules cannot be detected by the above techniques. Fluorescence-based methods would, in principle, allow \textit{in situ} characterization (i.e. with the  substrates used for cellular studies), as well as detection of small surface densities of adsorbed molecules. 

Using fluorescently-labelled molecules, a simple image provides relative information on molecular density through its intensity, but estimating the absolute value of the number of molecules requires additional and fragile calibrations \cite{Waters2014}. There are alternative approaches based on single molecule strategies, but they are adapted only for very low surface densities \cite{Verdaasdonk2014}. On the other hand, Fluorescence Fluctuation Spectroscopy (FFS) techniques \cite{Slaughter2010}, and more specifically Image Correlation Spectroscopy (ICS) which is suited to the characterization of immobile molecules, are intended for absolute quantification. These methods derive historically from Fluorescence Correlation Spectroscopy (FCS) and are based on the notion that signal originating from a submicrometric region within the sample, corresponding to the Point Spread Function (PSF) of the microscope, exhibits statistical fluctuations, since this PSF region does not always include the same number of molecules. The relative amplitude of these fluctuations provides an estimation of the average number of molecules in the PSF region and hence the concentration or density. Molecules cross the PSF either by spontaneous motion, as in the case of FCS, or by scanning the excitation laser (as in confocal microscopy), as in the case of ICS. ICS has been used to assess oligomerization\cite{Hennen2018}, or qualitatively detect the presence of aggregates through an increase of brightness or a corresponding decrease of number density \cite{Kolin2007, Kitamura2018}.

In fact, standard FFS techniques can only provide reliable quantitative information if all molecules have the same brightness (or if the distribution of brightness is known). Yet commercially available large proteins are rarely all labelled with the same number of fluorophores, so that they present a distribution of brightness. In this case, the number density of fluorescent entities (molecules or particles) estimated by conventional FFS is underestimated: indeed, part of the measured fluctuations would be due to variations in brightness and not only in number of entities, as usually assumed in FFS. More precisely, if the brightness distribution is characterized by a mean value, $\bar{\epsilon}$ and a standard deviation, $\sigma_{\epsilon}$, it can be shown (following \cite{Kolin2007,Mller2004}) that the mean number of entities in the PSF volume or area, as measured by FFS, is related to the true number of entities, $N$, by:
\begin{equation} \label{eq:N vs Nmeas}
N_{FFS} = \frac{N}{1+(\sigma_{\epsilon}/\bar{\epsilon})^2}
\end{equation}
Consequently, the wider the brightness distribution, the more pronounced the underestimation of the number density. The same bias affects the estimation of average brightness (which tends to be overestimated). Using only standard FFS methods, there is no way to evaluate this bias.

The present work aims at using photobleaching as a control parameter to measure accurately the density of surfaces coated with multiply-labelled entities. The combination of FFS and photobleaching has been proposed previously and used to estimate the degree of fluorescent labelling \cite{Delon2010}, the size of oligomers \cite{Ciccotosto2013,Paviolo2018} or the surface density of molecules \cite{DeMets2014}. However the practical implementations were either limited to the specific case of a Poisson distribution of brightness \cite{Delon2010,DeMets2014}, or failed to decipher the real parameters that could be deduced from the photobleaching decay \cite{Ciccotosto2013,Paviolo2018}. We present here a complete theoretical description of a method combining ICS to photobleaching, which is not specific to ICS and can be extended to other FFS techniques (for instance FCS, as in \cite{Delon2010}): we name it \emph{pbFFS} for \textit{photobleaching FFS}. Compared to previous work, we present a more in-depth theoretical derivation. We show that the measured brightness always decays linearly with photobleaching, whatever the distribution of fluorescent labels, and hence exactly two outputs can be extracted from this decay: (i) the brightness of a single fluorescent label, (ii) a factor depending on the mean and variance of the number of fluorescent labels per entity. We stress the fact that the presented method has the advantage to be \emph{calibration free}.\par
The pbFFS method is experimentally validated on substrates covered with monolayers of streptavidin molecules via a linker, used as a platform to build biomimetic surfaces step by step by adsorbing biotinylated molecules on top of it \cite{SefkowWerner2020, Dundas2013}. We applied pbFFS to streptavidin (SAv), fluorescently tagged with Alexa (SAv-Alex). Thanks to an additional assumption about the fluorescent labelling, we could bracket the mean number of fluorescent labels per SAv and, hence, estimate the absolute number density of SAv molecules covering the substrate. The resulting densities have been validated by independent spectroscopic ellipsometry measurements. We show that pbFFS is capable of measuring SAv surface densities that span over two orders of magnitude. Then, pbFFS was used to quantify the number of biotinylated fluorescent molecules (Atto labelled biotin, bAtto) that attach to a streptavidin base layer. In this case, the mass of adsorbed molecules was too low to be measured with accuracy by QCM-D or ellipsometry (since the mass of bAtto is only 1 kDa as compared to SAv which is 55 kDa). Interestingly, we also found by performing control photobleaching-FCS experiments that bAtto molecules are prone to aggregation in solutions, which potentially impacts the way they bind to the SAv base layer.\par

\section{Principles of photobleaching Fluctuation Fluorescence Spectroscopy}
\begin{figure}[ht!]
\centering
\includegraphics[width=0.9\textwidth]{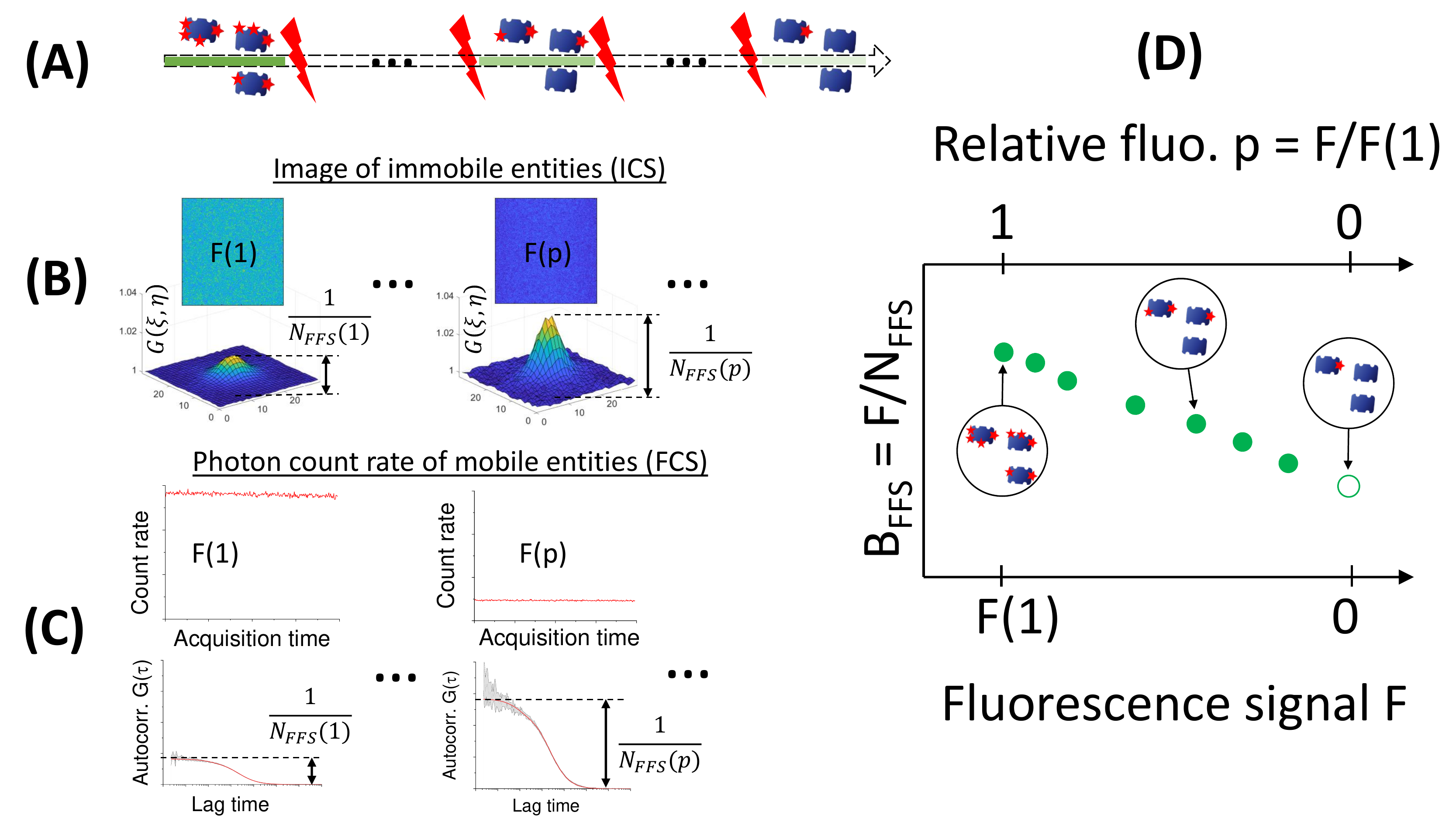}
\caption{Principle of a pbFFS experiment. Part (A) shows, along the dashed arrow, the photobleaching phases (red flashes) that alternate with the measurement phases (green sections of decreasing intensities). During the latter, the number of unbleached fluorescent labels borne by the entities decreases. The fluorescence signal, $F$, corresponds to the mean image intensity of immobilized fluorescent entities measured by ICS (B), or to the mean photon count rate of mobile fluorescent entities measured by FCS (C). The number of entities measured by FFS, $N_{FFS}$, also decreases as can be seen by the increased amplitude of the spatial autocorrelation in ICS, $G(\xi,\eta)$, or temporal autocorrelation in FCS, $G(\tau)$. Part (D) shows that the brightness measured during the successive phases, defined as $B_{FFS} = F/N_{FFS}$, plotted \emph{vs} the fluorescence signal, monotonously decreases upon photobleaching and tends towards that of a single fluorescent label.} 
\label{Acquisition method}
\end{figure}

As illustrated in Fig.~\ref{Acquisition method}A and D, a pbFFS experiment consists in alternating photobleaching phases, where a high laser power is sent into the sample to bleach a fraction of the fluorescent labels (e.g. Alexa) borne by the entities (e.g. streptavidin) and measurement phases at a reduced laser power. During the latter, the fluorescence signal, \textit{F}, the mean number of entities, $N_{FFS}$ and the brightness, defined as $B_{FFS} = F/N_{FFS}$, are measured. As more fluorescent labels are photobleached, the image intensity (in ICS, Fig.~\ref{Acquisition method}B) or the photon count rate (in FCS, Fig.~\ref{Acquisition method}C) decreases, while the autocorrelation amplitude increases, denoting a reduction of the number of entities $N_{FFS}$ in the PSF submicrometric region.

We consider a sample consisting of entities labelled with multiple fluorophores of identical brightness, $\epsilon$. The distribution of the number of these labels per entity before photobleaching is characterized by its mean and its standard deviation, named respectively $m$ and $\sigma$. It can be shown (see Section S1 of the Supporting Information (SI)) that the measured brightness $B_{FFS}$ is an affine function of the photobleaching stage $p$, defined as the fluorescence signal normalized to its initial value, i.e. $p = F/F(1)$:
\begin{equation} 
\label{eq: B final a}
B_{FFS}(p)=\epsilon(1+S_{\sigma m}p)
\end{equation}
where $S_{\sigma m}$ is the slope normalized by the single label brightness. It is related to the standard deviation and  mean value of the distribution according to:
\begin{equation} 
\label{eq: B final b}
S_{\sigma m}=\sigma^2/m +m-1
\end{equation}

The true total number of entities is then given by:
\begin{equation} \label{eq: N vs F m}
N = F(1)/m\epsilon
\end{equation}
The linear decay of the measured brightness \emph{versus} the fluorescence signal is valid for any distribution of fluorescent labels, assuming that all fluorescent labels have the same brightness, $\epsilon$ and a constant probability to bleach, whatever their number in the entities. A straightforward result of any pbFFS experiment is the single label brightness, $\epsilon$. This is an interesting output since, using Eq.~\ref{eq: N vs F m}, one can derive the total number of fluorescent labels, $Nm = F(1)/\epsilon$ . The exploitation of the slope, $S_{\sigma m}$, is more complex, because we see from Eq.~\ref{eq: B final b} that, in the general case, it is impossible to independently determine the mean value, $m$ and the standard deviation, $\sigma$, of the number of fluorescent labels per entity. However, specific fluorescent label distribution add constraints such that, in practice, the $m$ values are reduced to some range (For details see Section S2 in SI). This is the case for the SAv monolayer, as discussed in the next section.

\section{Implementation of the analysis method}

\subsection{Effect of background}

\begin{figure}[ht!]
\centering
\includegraphics[width=0.9\textwidth]{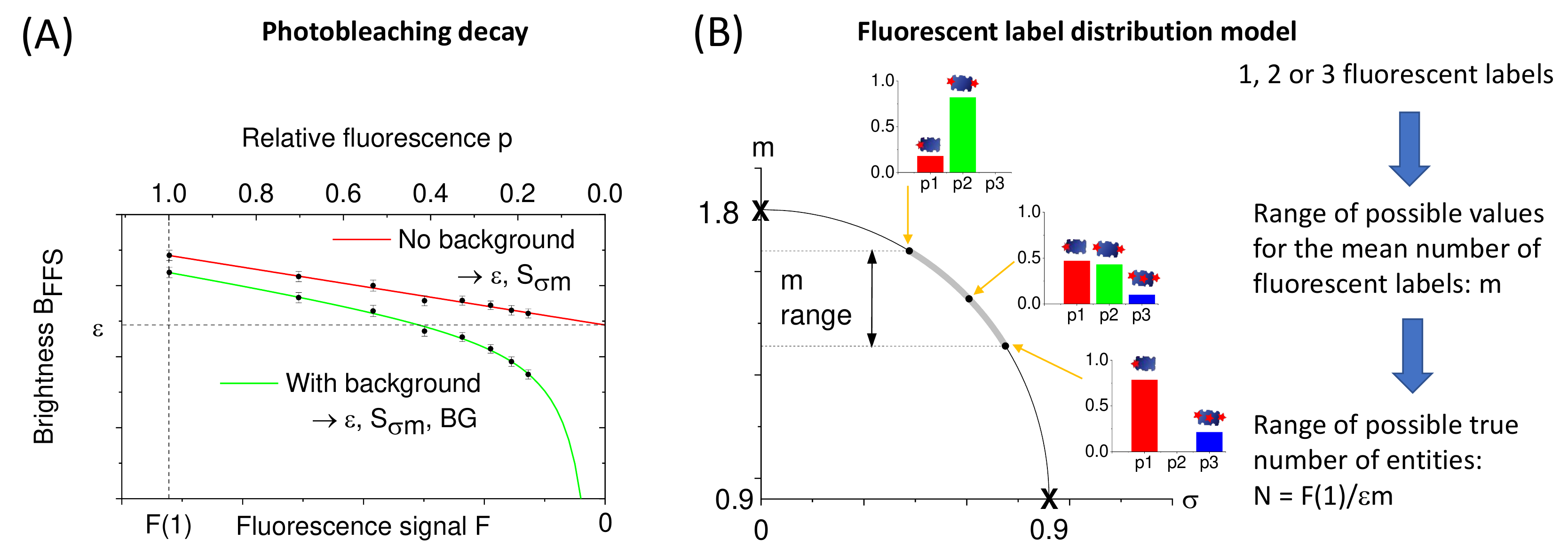}
\caption{Schematics of the implementation of the brightness decay analysis: (A) Upon photobleaching, the brightness, $B_{FFS}$, decreases linearly \textit{versus} the fluorescence signal, $F$ or $p$ (red solid line), thus providing the output parameters $\epsilon$ and $S_{\sigma m}$; in case of background (green solid line) an additional parameter, BG, can be estimated by fitting the decay (see text for details). (B) By combining the relation between $m$ and $\sigma$ corresponding to the measured value of $S_{\sigma m}$ and the constraints deriving from three non-nil probabilities ($p_1$, $p_2$ and $p_3$ for 1, 2 or 3 fluorescent labels), the possible $m$ values are restricted to a limited range; three examples of distributions of probabilities corresponding to the minimum, maximum and central values of $m$ are depicted for $S_{\sigma m}=0.8$. Finally, this analysis provides a value of the mean number of fluorescent labels, $m$, which combined with the initial fluorescence signal, $F(1)$ and the parameter $\epsilon$, leads to an estimation of the true number of entities, $N$.} 
\label{Analysis method}
\end{figure}

When performing pbFFS experiment we sometimes observe, both on surface and in solution, photobleaching decays that do not exhibit a linear behavior (as the red line in the schematic Fig.~\ref{Analysis method}A), but drop down to lower values of the brightness when the fluorescence signal gets small (green curve). This is due to an uncorrelated background, \textit{BG}, that contributes to the detected signal, thus making the relative fluctuations smaller and consequently the apparent number of entities larger, hence a lower brightness, as described in \cite{Brock1998}. It is nevertheless possible to incorporate the background as a free parameter in the photobleaching decay analysis by rewriting Eq. \ref{eq: B final a} as:
\begin{equation}
B_{FFS}(p)=\epsilon \left(1-\frac{r_{BG}}{p}\right)\left(1+S_{\sigma m}\frac{p-r_{BG}}{1-r_{BG}}\right)
\end{equation}

where $r_{BG} = BG/F(1)$ stands for the background normalized to the total initial signal (i.e. including the background itself). Correspondingly, $p$ becomes the relative total signal, including the background. The latter originates, either from light scattering (due to the glass substrate or to the walls of the PDMS microwells), or from some bulk fluorescence that would be not properly filtered out by the confocal detection. This effect appeared to be especially pronounced and difficult to mitigate in solutions. The fit of the photobleaching decay shown in Fig.~\ref{Analysis method}A thus gives the parameters $\epsilon$, $S_{\sigma m}$ (and $BG$, if relevant). \par

\subsection{Taking into account the fluorescent label distribution}

The degree of fluorescent labelling of SAv-Alex given by the manufacturer being 2, we take this number as an approximate value and consider that each SAv protein can bear 1, 2 or 3 Alexa labels, with probabilities $p_1$, $p_2$ and $p_3$. The key point is that this discrete distribution of probabilities has only 2 independent degrees of freedom (since, once the probabilities $p_1$ and $p_2$ are known, the last one, $p_3$, automatically follows). Additionally, we show in SI, Section S1, that, due to Eq. \ref{eq: B final b}, the $\sigma$, $m$ values are constrained along a circle. Combined with the properties of the distribution, this leads to a limited range of possible mean number of fluorescent labels. In our current situation, it can be shown (see SI, Section S2, Eq. S11 to S15 and Fig. S1) that if $S_{\sigma m}\leq1$, then $\frac{3}{3-S_{\sigma m}}\leq m\leq\frac{2}{2-S_{\sigma m}}$, while if $1<S_{\sigma m}\leq2$, then $\frac{3}{3-S_{\sigma m}}\leq m\leq\frac{6}{4-S_{\sigma m}}$ (we never measured $S_{\sigma m}>2$). Fig.~\ref{Analysis method}B schematically shows a few examples of distributions of probabilities $(p_1, p_2,p_3)$, which correspond to the value $S_{\sigma m}=0.8$ that is often found for SAv-Alex, spanning from the narrowest distribution (minimum value of $\sigma$), with the largest mean value, $m$, to the widest one, with the smallest $m$. Clearly, the latter distribution is unrealistic, as it would correspond to SAv proteins that never bear 2 fluorescent labels, while most of them bear 1 fluorescent label and some 3! It can be shown (SI, Section S2) that the condition to have a probability to bear 3 fluorescent labels smaller than that for 2 implies $m > \frac{1}{1-\frac{3}{8}S_{\sigma m}}$.

Interestingly, if we had hypothesised that SAv proteins could bear 2, 3 or 4 Alexa labels, this would lead to, either no solution when $S_{\sigma m}<1$ or, for the few cases where we measured $1\leq S_{\sigma m}\leq1.5$, to a vast majority of SAv molecules bearing 2 Alexa labels and almost none bearing 3 or 4, which is consistent with the model retained throughout this work (see  Section S2 of SI, Eq. S16). \par
Concerning the biotinylated fluorescent molecules, it is clear from the bAtto formula (given in the product specification) that they correspond to one single Atto dye. However, as already noted above, as a SAv protein of the base layer exposes from 1 to 3 free biotin binding sites, several bAtto molecules can colocalize on a single SAv protein. Therefore, we also assume that biotinylated entities can bear 1, 2 and 3 fluorescent labels.\par 
To conclude, from the analysis of the brightness decay, we obtain a range of values for $m$, and hence a range of values for the true number of fluorescent entities, $N=F(1)/m\epsilon$, which in turn leads to the final surface density given by $N/\pi {w_r}^2$.

\section{Experimental results}

\subsection{Streptavidin layers are densely packed}
Photobleaching ICS experiments have been performed on SAv layers with different percentages of fluorescently labelled SAv (SAv-Alex) of 1\%, 10\%, 50\% and 100\%. Since ICS usually performs better for relatively low surface densities, these dilutions aimed at testing the robustness of our method in a range of densities relevant for biomimetic surfaces. See Section S4 of SI about sample preparation, acquisition protocol and data analysis.

\begin{figure}[ht!]
\centering
\includegraphics[width=0.9\textwidth]{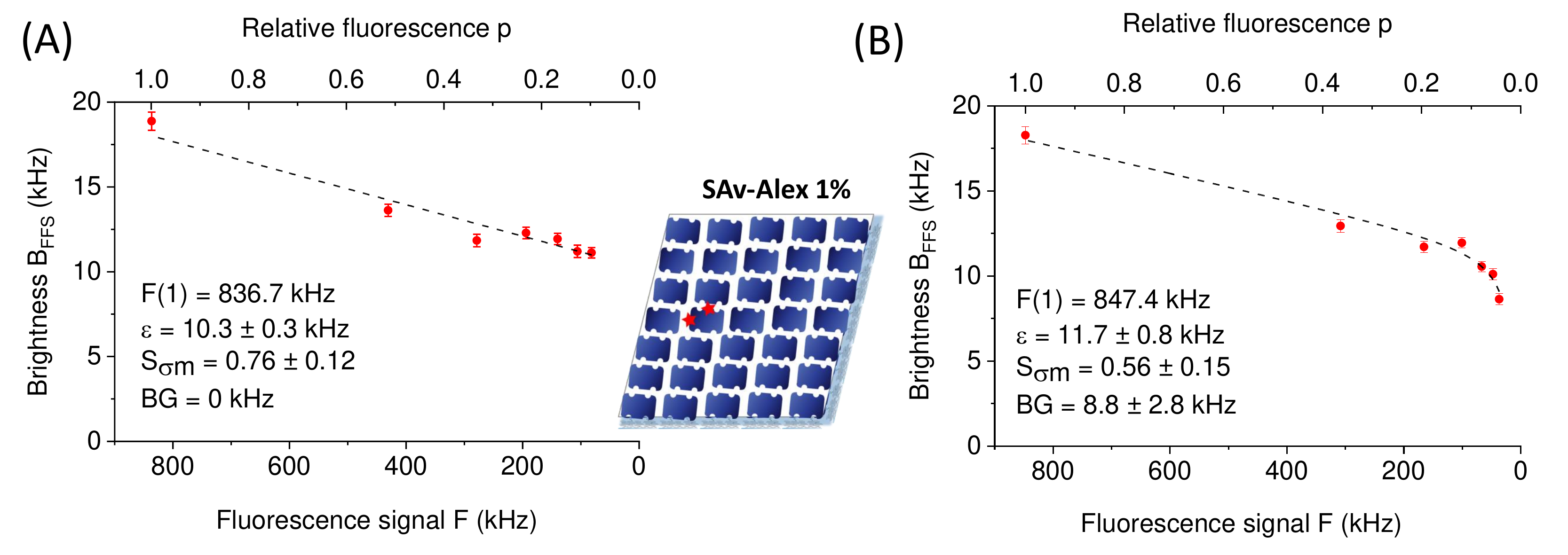}
\caption{Characteristic brightness decay \textit{versus} the fluorescence signal for a SAv layer prepared with 1\% dilution of SAv-Alex. (A) Example without background (i.e. fixed to 0), showing the linear decay and the fit (dashed black line) with the $\epsilon$ and $S_{\sigma m}$ outputs corresponding to the measured (red) points; the errors bars are the standard errors of the mean calculated from $8\times8$ sub-images; (B) In presence of background, the corresponding parameter, $BG$, can be fitted (it represents about 1\% the initial fluorescence signal, $F(1)$), while the other outputs, $\epsilon$ and $S_{\sigma m}$, are consistent with the results of the background-free case.}
\label{SAv-Alex1-A-B}
\end{figure}

We see in Fig.~\ref{SAv-Alex1-A-B}A a plot of a characteristic brightness decay \textit{versus} the fluorescence signal obtained for 1\% dilution of SAv-Alex. The parameters estimated from the fit, $\epsilon=10.3$ kHz and $S_{\sigma m}=0.76$, correspond to an initial brightness, given by $B_{FFS}(1)=\epsilon(1+S_{\sigma m})$, that is about 1.8 times larger than the single fluorescent label brightness. Consequently, if the distribution of the number of fluorescent labels per molecule was single-valued, one would be close to the situation of 2 Alexa dyes per SAv, which corresponds to the manufacturer specifications. However, using the theory presented in Section S2 of SI, we can estimate the range of $m$ values compatible with both the possible numbers of fluorescent labels and the estimated value of $S_{\sigma m}$ at $[1.40,1.61]$ (see SI, Section S2, Fig. S1 and Eq. S14). This corresponds to a surface density of fluorescently labelled SAv, at 1\% dilution, in the range of 316-363 molecules per $\mu m^2$ (using $w_r$= 0.22~$\mu$m for this data set). We also show, in Fig.~\ref{SAv-Alex1-A-B}B, another result obtained with the same dilution of SAv-Alex, but exemplifying the consequence of a background on the brightness decay. Although the background accounts only for about 1\% of the total initial signal, the deformation from linearity of the brightness decay is very pronounced at the end of the process. However, the decay can still be well fitted using one additional parameter, $BG$, while the other parameters remain consistent with their values estimated without background.\par

\begin{figure}[ht!]
\centering
\includegraphics[width=0.45\textwidth]{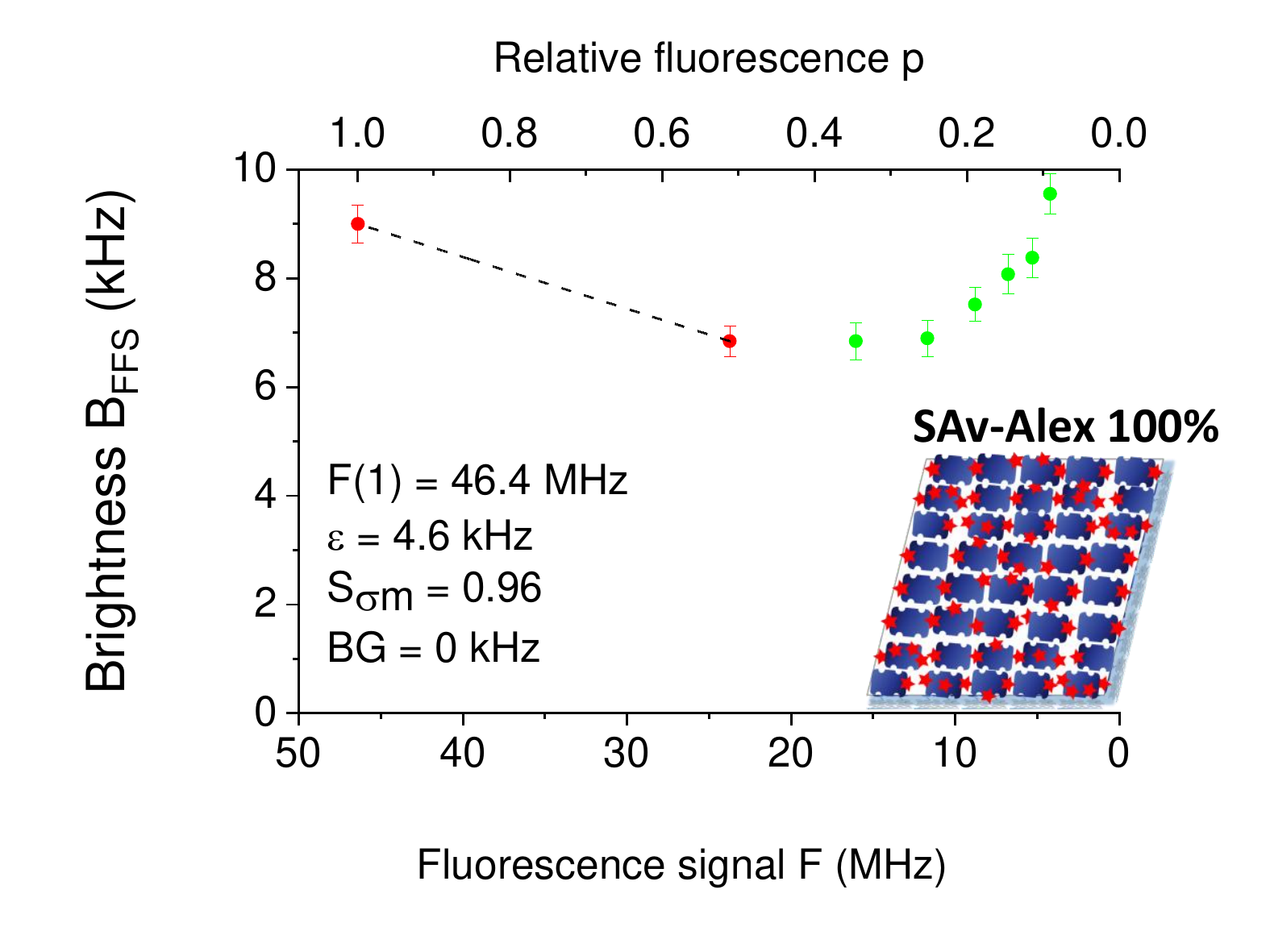}
\caption{Example of the dramatic consequence, for the brightness decay, of the fluorescence quenching between Alexa555 fluorophores of a streptavidin layer prepared with 100\% of SAv-Alex. Note that the single fluorescent label brightness, $\epsilon$, is more than twice smaller than its value at high dilution (see Fig.~\ref{SAv-Alex1-A-B}), while the $S_{\sigma m}$ parameter, estimated from the 2 first points, stays consistent.}
\label{SAv-Alexa100-quenching}
\end{figure}

A physically different situation occurs at high surface densities of fluorescently labelled SAv, as shown in Fig.~\ref{SAv-Alexa100-quenching} for a layer containing 100\% of SAv-Alex. This curve exhibits a dramatic brightness recovery after an initial decay. By fitting the first part of the brightness variation, assumed to be linear, we estimated the $S_{\sigma m}$ parameter at a value, 0.96, reasonably close to the one measured at 1\% dilution. Conversely, the single fluorescent label brightness, $\epsilon=4.6$ kHz, is found to be more than 2 times lower than the value measured at high dilutions of SAv-Alex (see Fig.~\ref{SAv-Alex1-A-B}). We suggest that fluorescence quenching between identical fluorophores can explain this behavior: at 100\% concentration, the mean distance between the Alexa555 dyes is of the order of the size of the streptavidin molecules, that is $\approx$ 5 nm. At such short distances, the reduction of the fluorescence quantum yield can be very pronounced \cite{Bae2021}. When photobleaching occurs, the mean distance between intact fluorescent labels increases, thereby reducing self-quenching so that the brightness tends towards its normal value. Because this is not within the scope of the present work, we did not study further the exact shape of the brightness curve, but we tentatively made use of the 100\% concentration results to estimate the surface density of SAv. Using the waist size found for this 100\% SAv-Alex case, $w_r = 0.25 \mu m$ and $m=1.7$ (middle of the range corresponding to $S_{\sigma m}=0.96$), we estimate 30330 streptavidin molecules per $\mu m^2$. First, this number is close to 100 times the surface density measured at 1\% dilution. Second, it corresponds to a mean distance of 5.7 nm between streptavidins, which is consistent with the assumption of a densely packed layer, given the size of a streptavidin molecule\cite{Hendrickson1989}. Therefore, although based solely on the first 2 points, the results obtained for 100\% of fluorescently labelled SAv are compatible with the lower-density case. Experiments have also been performed with 10\% and 50\% dilutions, the results of which are reported and synthesized in Table \ref{tab:Summary of results}. Overall, our SAv-Alex results are consistent with each other, with a clear trend towards more fluorescent quenching as the surface density of SAv-Alex increases.

\subsection{Streptavidin in solution has the same fluorescent label distribution }
\begin{figure}[ht!]
\centering
\includegraphics[width=0.9\textwidth]{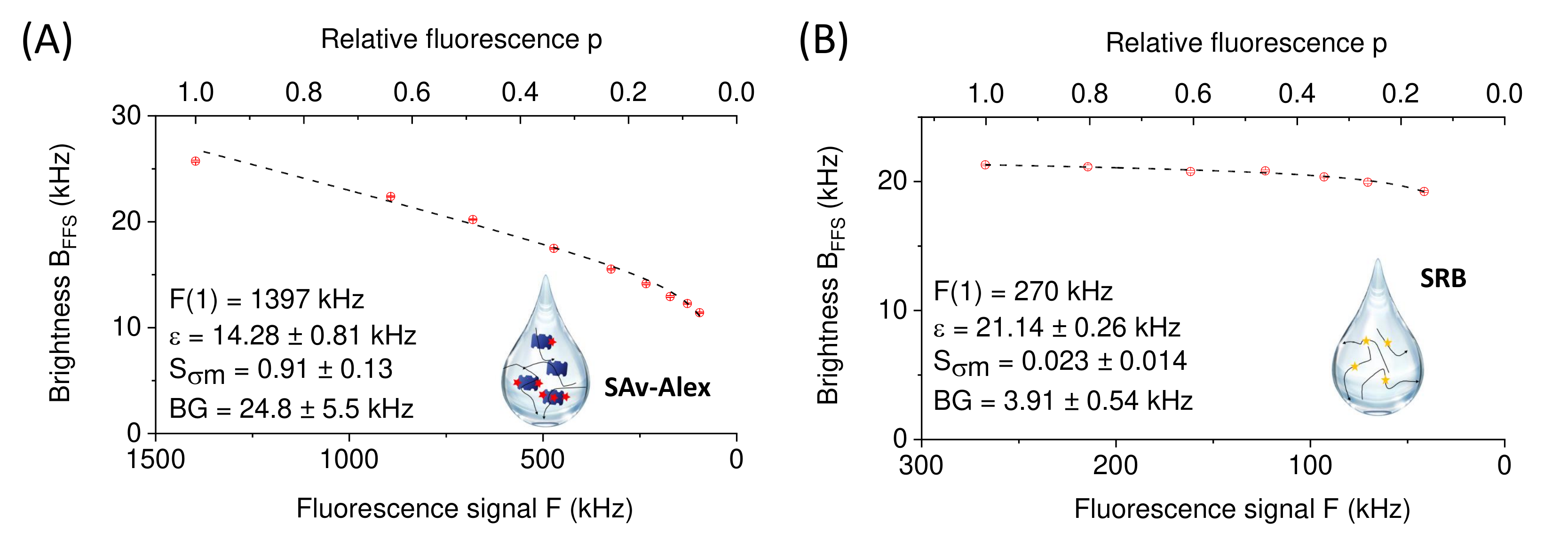}
\caption{Examples of brightness decays measured by FCS, for solutions confined in PDMS microwells. (A) SAv-Alex molecules, where the normalised slope, $S_{\sigma m}$, is close to its value found on surfaces (see Fig.~\ref{SAv-Alex1-A-B}); (B) SRB in HEPES, clearly showing an almost constant brightness, in agreement with the expected behavior of a single dye molecule. For both cases a background of up to 2\% is estimated. Note that the error bars are smaller than the point size.}
\label{Solution-A-B}
\end{figure}

In order to get an independent estimation of the fluorescent label distribution, we also applied our method to SAv-Alex molecules freely diffusing in solution. We found a value of $S_{\sigma m}$ very consistent with the estimations on surfaces (around 0.8), as illustrated in Fig.~\ref{Solution-A-B}A. As a control case, Fig.~\ref{Solution-A-B}B shows that Sulforhodamine B sodium salt (SRB) solutions lead to a constant brightness when photobleaching, as expected for a single dye molecule. 

\subsection{Biotinylated molecules bind to about 10\% of the streptavidin base layer }
\begin{figure}[ht!]
\centering
\includegraphics[width=0.9\textwidth]{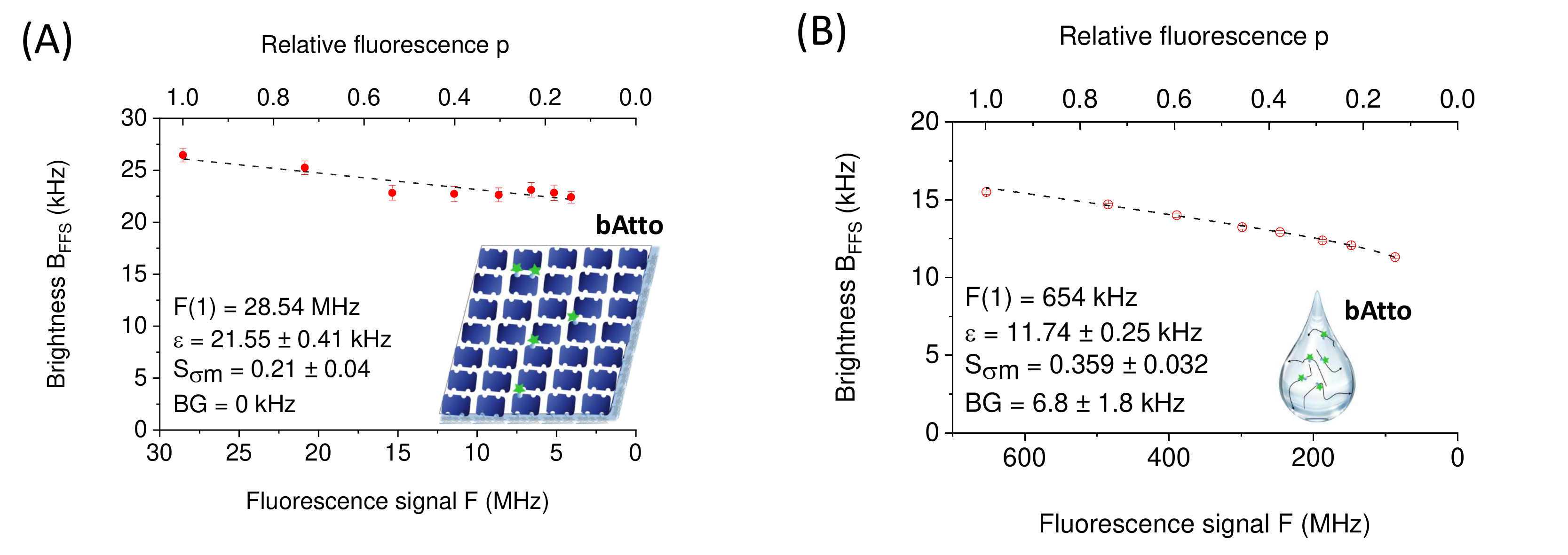}
\caption{Brightness decay of biotinylated fluorescent molecules (bAtto), measured on surface and in solution. (A) ICS measurements for bAtto bound to a streptavidin base layer, exemplifying the lower normalised slope value, $S_{\sigma m}=0.21$, compared to SAv-Alex (the fact that bAtto exhibits a single fluorescent label brightness different from that of Alexa555 labelled SAv is only due to the photophysical properties of these dyes); (B) FCS measurements for bAtto freely diffusing in solution. Note that the error bars are smaller than the point size.}
\label{bAtto-solution-surface}
\end{figure}

Next we investigated the surface density of biotinylated molecules deposited on top of SAv layers, in order to show the potential of pbFFS as a characterization tool for developing biomimetic surfaces. Fluorescent bAtto molecules bound on a streptavidin base layer were observed. The brightness decay, as shown in Fig.~\ref{bAtto-solution-surface}A, exhibits a normalized slope, $S_{\sigma m}=0.21$, which is significantly smaller than the one observed for a SAv-Alex layer. This means that this case is close to the one of a single fluorescent label per labelled SAv. Assuming the same type of distribution (i.e. over 1, 2 or 3 fluorescent labels), the range of $m$ values that corresponds to $S_{\sigma m}=0.21$ is estimated to $[1.08,1.12]$ (see SI, Fig. S1 and Eq. S14). Two quantities can be estimated from these measurements: the surface density of bAtto molecules given by $F(1)/\epsilon \pi {w_r}^2$, which is 3708 molecules per $\mu m^2$ and the density of \emph{fluorescently labelled} SAv given by $F(1)/(m\epsilon \pi {w_r}^2)$ with $m=1.1$, center of the range estimated above, which is approximately 3370$/ \mu m^2$. Since we found previously $\sim$30330 SAv per $\mu m^2$ in the base layer, this implies that about 1 in 10 SAv molecules can carry a bAtto molecule. Note that the areal mass density reported, from different experiments, in Table \ref{tab:Summary of results} is that of bAtto molecules. 

Each SAv has 4 biotin binding sites: at least one of these is used to bind to the PLL-g-PEGbiotin base layer, so that the \textit{a priori} number of available sites for bAtto is between 0 and 3. Our data indicate that many SAv molecules could not bind any bAtto, presumably due to steric hindrance. More precisely, using $m \sim$1.1, we can deduce that among the \emph{fluorescently labelled} SAv, about 90\% of SAv bound only one bAtto (see Eq. S12 in SI), leaving 10\% that carry more than one bAtto. Two mechanisms can lead to this distribution: either bAtto bind to SAv as single molecules, in which case a few SAv have more than one occupied biotin-binding pocket, or bAtto preexist as dimers, trimers, etc. in the solution used for incubation, in which case, SAv may present only one pocket occupied by a bAtto complex. Indeed, the fact that Atto dyes are moderately hydrophilic \cite{ZanettiDomingues2013} could favor aggregation.

To assess bAtto aggregation in solution, we performed pbFFS experiments with bAtto solutions (FCS), as reported in the next paragraph.\par

\subsection{Biotinylated molecules slightly aggregate in solution}
We show in Fig.~\ref{bAtto-solution-surface}B an example of a photobleaching experiments with bAtto in solution. Globally, we recovered values of the normalised slope close to those found on surface (i.e. $S_{\sigma m} \approx 0.2 - 0.3$), which is clearly distinct from the single molecule case of SRB in solution shown in Fig.~\ref{Solution-A-B}B. This range of values for $S_{\sigma m}$ corresponds (see SI, Fig. S1) to a mean number of fluorescent labels per diffusing entity, $m$, between 1 and 1.2, as summarized in Table \ref{tab:Summary of results}. Since one bAtto molecule definitively corresponds to a single Atto dye, we conclude that unspecific aggregation occurs between bAtto molecules in solution, despite the moderate concentration (< $\mu$M). As a consequence, although the concentrations used to incubate the SAv base layer with bAtto are much lower than those used for FCS experiments (which should mitigate the aggregation trend), we cannot exclude that each SAv has only one bAtto binding site occupied by an aggregate. In this case, the number of available biotin binding sites on the SAv layer is not that of the number of bound bAtto molecules (i.e. 3708 / $\mu m^2$) but rather that of the number of  SAv molecules labelled with bAtto (i.e. 3370 / $\mu m^2$).\par
Note that the difference in brightness between  bAtto in solution and on a SAv layer, shown in Fig.~\ref{bAtto-solution-surface}, is not relevant, as the ICS and FCS experiments are performed on surface and in bulk, respectively (with different microscopy setups). The same remark stands for SAv results shown in Fig.~\ref{Solution-A-B}A and Fig.~\ref{SAv-Alex1-A-B}.

\begin{table}[htb]
    \centering
    \begin{tabular}{c c c c }
        \textbf{Molecule} & \textbf{Sample} & \textbf{Fluo. Label.} & \textbf{Density ($ng/cm^2$)}\\
        SAv-Alex & Surface 100\% & 1.79 $\pm$ 0.39 & 239 $\pm$ 34\\
        SAv-Alex & Surface 50\% & 1.22 $\pm$ 0.05 & 303 $\pm$ 50\\
        SAv-Alex & Surface 10\% & 1.94 $\pm$ 0.22 & 293 $\pm$ 69\\
        SAv-Alex & Surface 1\% & 1.74 $\pm$ 0.36 & 284 $\pm$ 70\\
        SAv-Alex & Solution & 1.58 $\pm$ 0.18 & n/a\\
        bAtto & Surface & 1.13 $\pm$ 0.09 & 0.82 $\pm$ 0.23\\
        bAtto & Solution & 1.125 $\pm$ 0.057 & n/a\\    
        SRB & Solution & 1.013 $\pm$ 0.011 & n/a  
    \end{tabular}
    \caption{Degree of fluorescent labelling ($m$ values) and areal mass density (when applicable) measured with pbFFS for different molecules, on surface and in solution. For SAv-Alex molecules diluted at 50, 10 and 1\%, the estimated areal mass density is linearly extrapolated to 100\% SAv. Values are given as mean $\pm$ SD}
    \label{tab:Summary of results}
\end{table}

\subsection{pbFFS results agree with spectroscopic ellipsometry}

We present in Table \ref{tab:Summary of results} the degree of fluorescent labelling ($m$) corresponding to different molecules, SAv-Alex, bAtto and SRB, in solution and on surfaces. For each case, the lower and upper values of the $m$ ranges estimated for the different experiments were pooled altogether to give a mean and a standard deviation (SD) of $m$. When applicable, these values were then used to derive the areal mass density of SAv molecules. Although the photobleaching decay curves of 100\% SAv-Alex have a limited number of exploitable points (due to the above mentioned quenching effect), the corresponding pbFFS estimation of the density  is in fairly good agreement with the spectroscopic ellipsometry result that gives 246 $\pm$ 36 $ng/cm^2$ \cite{SEdata}. The linear extrapolation to 100\% SAv of the SAv-Alex molecules diluted at 1, 10 and 50\% tends to show a higher areal mass density, but this effect may be related to the uncertainty of dilutions. We also stress the fact that, although the degree of labelling found for 50\% dilution differs from the other cases, the corresponding areal mass density stays consistent with the other results.

Concerning bAtto, the reported areal mass density of 0.82 $\pm$ 0.23 $ng/cm^2$ is that of the total number of bAtto molecules bound to the SAv base layer. This value holds whatever the distribution of bAtto on SAv molecules. Note that such low density could not be measured by ellipsometry, so that no comparison can be made.

\section{Discussion}

We have presented a method, named pbFFS, which aims at estimating the number density of entities of non uniform brightness. We have shown its potential for more reliable quantification of surface density in the context of biomimetic layer characterization. Hereafter, we discuss two limiting situations.

In this work, we have measured surface densities spanning two orders of magnitude, from $\sim$300 to $\sim$30000 molecules per $\mu$m$^2$. This latter value is probably close to the highest density that can be measured using ICS. Indeed, at high concentration/density of molecules, the relative intensity fluctuations and thus, the autocorrelation amplitude, are very small and may be hidden behind unwanted variations (non-uniformity of the imaging system or the sample). Therefore, it is quite important to properly flatten images, check stability of solutions, etc. High densities can also induce some detrimental photophysical effects, such as quenching. However, we experimentally showed that this does not prevent correct results, provided the biased data points are taken off from the analysis.

The streptavidin molecules studied here were labelled by 1 to 3 fluorophores, so the degree of fluorescent labelling was rather low. One may wonder whether pbFFS would also be suitable for studying entities with a higher degree of fluorescent labelling (several tens of fluorophores or more). In this case, the photobleaching process starts with entities that are initially very bright and must be carried out till only single fluorescent labels are left, which requires a high dynamic range of detection. Using fluorescent nanospheres we could check that, in principle, our framework also applies in such situations, as shown in SI, Section S3.

Besides number density quantification, some applications may benefit from the ability of pbFFS to provide information on the degree of fluorescent labelling (or number of fluorophores per entity). This would be the case of oligomerization studies.
Oligomerization is an ubiquitous phenomenon that plays an important role in numerous biological processes. The size of oligomers is usually estimated using fluorescence fluctuation methods (FCS, ICS) by comparing the measured brightness to the one found on a sample containing only monomers, in the same experimental conditions. Such methods suffer from two drawbacks: first, a reference sample with only monomers may not be easily available; second, if all the oligomers do not have the same size, the result is biased resulting in an overestimation of the oligomer size. More advanced techniques have been proposed, such as SpIDA\cite{Godin2015}, FIF \cite{Stoneman2019} or eN\&B\cite{Cutrale2019} but still rely on the monomer brightness. The pbFFS method does not require a separate measurement on monomers, since the monomer brightness ($\epsilon$) is provided by the fit: it is \emph{self-calibrated}, which is a significant advantage. Moreover, some information on the distribution of oligomer size can be obtained, although, as shown in this work, the exact distribution can only be resolved if the number of degrees of freedom is reduced by additional assumptions (\textit{e.g.} only few possible oligomers). This strategy would work not only on fixed samples, but also in living cells. Photobleaching has been proposed in live cell, outside of the oligomerization context, to retrieve molecular brightness and to turn confocal images into concentration maps \cite{Zhang2021}. However this method cannot be used in cases when the studied protein forms homo-oligomers, contrarily to the present work.

Finally, we point out that, while our method contrasts with standard semi-quantitative fluorescence approaches that cannot deal with a distribution of brightness, it can be compared with some single molecules techniques based on stepwise photobleaching\cite{Verdaasdonk2014, Wang2015}. For instance, Madl et al. performed a combination of photobleaching and brightness analysis to measure the subunit composition of membrane proteins\cite{Madl2010}. In a recent work, Stein et al. have combined FCS with single-molecule localization microscopy to provide the number of docking strands in spatially well-separated origami nanostructures \cite{Stein2019}. However, single molecules techniques obviously require extremely low surface density of immobilized molecules, while pbFFS works in much higher density regimes, which are more relevant for most biomimetic and biological samples.

\section{Conclusion}

In this paper we presented a method, combining fluctuations analysis and photobleaching, that aims at characterizing molecules or particles fluorescently labelled with an unknown distribution of fluorophores. pbFFS can be used both on surfaces or in volume, with immobilized or moving molecules (either in flow or freely diffusing), observed in confocal or TIRF microscopy performed in photon counting mode. It provides the single fluorescent label brightness, as well as a parameter depending on the mean and variance of the distribution of fluorescent labels. If additional assumptions can be used to restrict the number of degrees of freedom of this distribution, the degree of fluorescent labelling and an unbiased value of the concentration or density can be deduced.

The pbFFS method has been demonstrated on a SAv base layer of biomimetic samples to estimate the surface density of SAv, and its propensity to bind a top layer of bAtto molecules. For the base layer density, the density estimated with pbFFS is in agreement with ellipsometry measurements. However, compared to this latter technique or QCM-D, pbFFS has the advantage of allowing \textit{in situ} characterization, since it does not require dedicated substrates, and allows the quantification of low mass of adsorbed molecules (as in the case of bAtto), out of reach of the other techniques, thanks to the intrinsic sensitivity of fluorescence measurements.

We believe pbFFS can provide a powerful framework to attain more reliable fluorescence fluctuations analysis: indeed, standard FFS techniques are completely unable to assess whether their results are biased by a dispersion of brightness values (or number of fluorescent labels), whereas, using pbFFS, the linear decay of the brightness during photobleaching is a simple checkpoint, which contributes to making the measurements more trustworthy.

In addition to number density estimations, the capability of pbFFS to evaluate the number of fluorophores per entity should make it particularly useful for oligomerization studies, as estimating protein oligomerization is essential to understand numerous cellular functions.

\begin{acknowledgement}
We acknowledge D. Centanni (LIPhy) for the microwell preparation and the IAB facility for the confocal microscope.  We thank M. Balland (LIPhy) and O. Destaing, A. Grichine (IAB) for stimulating discussions. We acknowledge E. Castro Ramirez for her work with PLL-g-PEGb. The project is funded by ANR GlyCON ANR-19-CE13-0031-01.

\end{acknowledgement}

\begin{suppinfo}

Section S1 - Theoretical derivation of the brightness decay and exploitation of the measurements; Section S2 - Theoretical derivation of occupancy probabilities and $m$ value ranges; Section S3 - Results and analysis of experiments with 20~nm fluorescent beads; Section S4 - Materials and methods (PDF).

\end{suppinfo}

\providecommand{\latin}[1]{#1}
\makeatletter
\providecommand{\doi}
  {\begingroup\let\do\@makeother\dospecials
  \catcode`\{=1 \catcode`\}=2 \doi@aux}
\providecommand{\doi@aux}[1]{\endgroup\texttt{#1}}
\makeatother
\providecommand*\mcitethebibliography{\thebibliography}
\csname @ifundefined\endcsname{endmcitethebibliography}
  {\let\endmcitethebibliography\endthebibliography}{}

\end{document}


\textbf{Summary:} {Theoretical derivation of the brightness decay and exploitation of the measurements (S1); Theoretical derivation of occupancy probabilities and $m$ value ranges (S2); Results and analysis of experiments with 20~nm fluorescent beads (S3); Materials and methods (S4).}

\section{S1. Theoretical derivation of the brightness decay and exploitation of the measurements}
\subsection{Theoretical derivation of the brightness decay}
In the following, we derive the rigorous expression of the brightness $B_{FFS}$ as a function of the photobleaching stage (defined as the fraction of remaining fluorescence signal), assuming all fluorophores are independent and have an equal probability to bleach. Although we present the formalism of pbFFS for two-dimensional ICS, the same results apply to mobile molecules and temporal techniques, \textit {e.g.} FCS in solution, Raster ICS \cite{Kolin2007}, etc.

Let us consider $I(x,y)$ the image intensity at pixel $x,y$. The fluorescence fluctuations, defined as $\delta I(x,y)=I(x,y)-\langle I(x,y)\rangle$ (where the averaging is performed over the image field), are analyzed using the autocorrelation function:
\begin{equation}
G(\xi,\eta)=\frac{\langle \delta I(x,y)\delta I(x+\xi,y+\eta) \rangle}{\langle I(x,y) \rangle ^2}
\end{equation}

In spatial ICS, the fluorescent entities of interest are immobile, so that the autocorrelation is only related to the optical resolution of the microscope (described by the PSF) and is very well approximated by a Gaussian \cite{Petersen1993}:
\begin{equation} \label{eq: G}
G(\xi,\eta)=\frac{1}{N_{FFS}} \exp \left(-\frac{\xi^2+\eta^2}{w_r^2}\right)
\end{equation}
where $N_{FFS}$ is an apparent mean number of entities in the PSF area of radius $w_r$. As already pointed out in the Introduction of the primary manuscript (Eq. 1), when the observed species are not equally bright, $N_{FFS}$ is smaller than $N$, the real number of all the entities. In the case of a mixture of species of different brightness, it can be shown that the FFS techniques leads to  \cite{Kolin2007,Mller2004}:
\begin{equation}
N_{FFS}=\frac{(\sum \epsilon_i N_i)^2}{\sum \epsilon_i^2 N_i}
\end{equation}
where $\epsilon_i$ is the brightness of the species $i$ (that is the fluorescence signal of a single entity of the species $i$) and $N_i$, the average number of entities of the species $i$. Note that the true total number of entities is given by $N = \sum N_i$. The key point is that the contributions are weighted by the square of the brightness, leading to an underestimation of the total number of fluorescent entities, when all entities are not equally bright (it's the tree that hides the forest). 

Here, we consider that each entity carries several identical fluorescent labels. In the forthcoming derivation we shall assume that fluorescence quenching can be neglected, so that the brightness of a single fluorescent label is constant whatever their number in the entities. In this case, the brightness of an entity carrying $n$ fluorescent labels is $\epsilon_n=n\epsilon$ where $\epsilon$ is the brightness of a single fluorescent label, and the fluorescence signal reads ($N_n$ being the number of entities that bear exactly $n$ fluorescent labels):
\begin{equation} \label{eq: F}
F=\sum n\epsilon N_n
\end{equation}

The overall brightness is thus given by:
\begin{equation} \label{eq: B}
B_{FFS}=\epsilon\frac{\sum n^2 N_n}{\sum n N_n}
\end{equation}

To describe the photobleaching effect, we present a derivation based on the same hypothesis as in the work by Ciccotosto \textit{et al.}\cite{Ciccotosto2013}. However, we generalize the formalism in order to provide a simple theoretical expression of the brightness for any initial distribution of fluorescent labels. At a given point during photobleaching, we assume that any fluorophore has the same probability not to be bleached, given by $p$. This implies that the initial number of fluorescent labels, $n$, which appears in Eq.~\ref{eq: F} and \ref{eq: B} has to be replaced by the mean number of unbleached labels, which is simply given by $np$. Therefore the remaining fluorescence signal reads:
\begin{equation} 
\label{eq: F bleached}
F(p)=\epsilon \sum np N_n=\epsilon pmN
\end{equation}

where $m=\frac{1}{N}\sum n N_n$ is the average initial number of fluorescent labels per entity, computed over all fluorescent entities (this can be called \emph{degree of fluorescent labelling}). Note that $p$ is nothing but the fluorescence signal normalized to its initial value, before photobleaching has started, i.e. $p=F(p)/F(1)$. In order to also modify Eq.~\ref{eq: B} and make it valid all along photobleaching, we need to replace $n^2$ by the second order moment of the distribution at the fluorescence stage $p$. Since any fluorescent label can only be in two states, bleached or unbleached with respective probabilities $(1-p)$ and $p$, this quantity results from the binomial distribution and equates $np(1-p) + n^2p^2$. As a consequence, Eq.~\ref{eq: B} becomes:
\begin{equation} 
\label{eq: B intermediate}
B_{FFS}(p)=\epsilon\frac{\sum [np(1-p) + n^2p^2] N_n}{pmN}
\end{equation}

We see that the numerator of Eq.~\ref{eq: B intermediate} reveals, in addition to the mean value, $m$, of the initial number of fluorescent labels per entity, the mean value of its square, $\frac{1}{N}\sum n^2 N_n$ that can be written ${\sigma}^2+m^2$, where $\sigma$ is the standard deviation of the initial number of fluorescent labels per entity. Consequently, after a few simplifications, Eq.~\ref{eq: B intermediate} can be written again:
\begin{equation} 
\label{eq: B final a}
B_{FFS}(p)=\epsilon(1+S_{\sigma m}p)
\end{equation}

where
\begin{equation} 
\label{eq: B final b}
S_{\sigma m}=\sigma^2/m +m-1
\end{equation}

Hence the measured brightness $B_{FFS}$ is an affine function of the photobleaching stage $p$: the single fluorescent label brightness $\epsilon$ is its intercept at $p=0$ (note that this is an immediate output of pbFFS, obtained without any assumption) and $S_{\sigma m}$ is its slope normalized by the single fluorescent label brightness. Let us rewrite here the equation stating the initial fluorescence signal:
\begin{equation} \label{eq: F ini}
F(1)=\epsilon mN
\end{equation}
Eq.~\ref{eq: B final a}, \ref{eq: B final b} and \ref{eq: F ini} are the core relations around which all our reasoning is based. Note that studying the variation of the number of entities, instead of the brightness, \textit{versus} the fluorescence signal, would be completely equivalent, as $B_{FFS}$ and $N_{FFS}$ are related through $F=B_{FFS}N_{FFS}$ (we drop out the $_{FFS}$ subscript in $F$ because the fluorescence signal is not biased by fluctuation measurements). We will mostly discuss the brightness because, in absence of background, it always decays as a straight line when plotted as a function of the fluorescence signal (whatever the initial distribution of fluorescent labels), which is quite convenient for visual inspection of the experimental results. At this stage, it is interesting to make a few remarks about the pbFFS approach.

\begin{itemize} 
\item First, we emphasize the fact that the parameters $\sigma$ and $m$ appearing in Eq.~\ref{eq: B final b} and \ref{eq: F ini} characterize the \emph{initial} distribution of brightness. Of course this distribution varies during photobleaching (it can be shown that it always converges towards a Poisson one), but the slope, $S_{\sigma m}$, depends only on the initial distribution.

\item Although $B_{FFS}(p)$ is an affine function of $p$ whatever the initial brightness distribution, a very peculiar situation is that of a mixture of entities bearing either no fluorescent label, or exactly one. This leads to $\sigma^2=m(1-m)$, hence $S_{\sigma m}=0$ and the measured brightness, $B_{FFS}(p)$, is constantly equal to $\epsilon$, independently of the photobleaching stage. Indeed, all visible entities can only have the brightness of single labels.

\item Another notable case is an ensemble of entities that initially bear the same number of labels, say an integer $m$ (this is a single-valued distribution, with $\sigma=0$). This would lead to a measured brightness, $B_{FFS}(p)$, that linearly varies from $m\epsilon$ to $\epsilon$.

\item In all cases, when the fluorescence signal vanishes ($p\rightarrow 0$), the measured brightness tends towards the one of a single label ($B_{FFS}(0)\rightarrow\epsilon$), since the only entities that remain visible are those bearing one single label. Consequently, one also obtains the total number of fluorophores that is just given by $mN = F(1)/\epsilon$.

\item Finally, it is worth to notice that in case of an unknown proportion of dark entities, there is obviously no way to assess it, and thus, to quantify the true total number of entities.
\end{itemize} 

\subsection{Exploitation of the measurements}
\begin{figure}[ht!]
\centering
\includegraphics[width=0.45\textwidth]{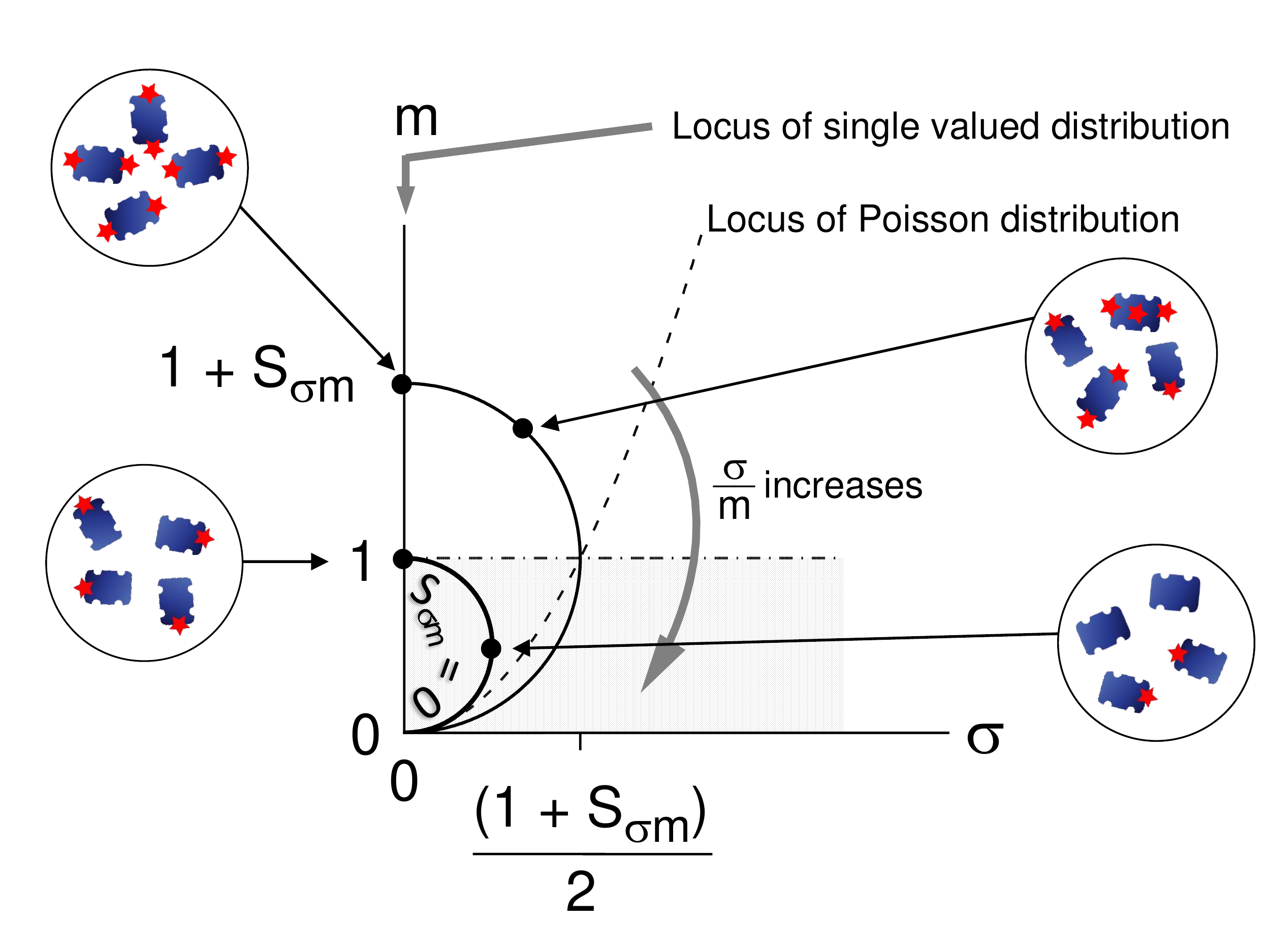}
\caption{Geometrical representation of the relation between the mean, $m$, and the standard deviation, $\sigma$, of the \emph{initial} number of fluorescent labels per entity. For a given measured slope, $S_{\sigma m}$, the support of the $(\sigma,m)$ points is a half circle of diameter $1+S_{\sigma m}$ that always crosses the $(0,0)$ point; the coefficient of variation of the distribution of the number of fluorescent labels, $\frac{\sigma}{m}$, continuously increases from the top to the base of the half-circle. Single-valued distributions correspond to discrete points located on the vertical axis ($\sigma =0$), as exemplified for 1 and 2 fluorescent labels. The smallest circle of diameter 1 ($S_{\sigma m} = 0$) corresponds to a mixture of entities bearing either no fluorescent label or exactly one, as depicted at the bottom right. If all molecules bear at least one fluorescent label, for instance a mixture of 1, 2 and 3 fluorescent labels, the $(\sigma,m)$ solutions are located in the upper half space above the dashed-dotted line $m= 1$. Note that in the case of a Poisson distribution, any given slope $S_{\sigma m}$ corresponds to a unique point $(\sigma,m)$ located at the intersection of the $m=\sigma^2$ curve (dashed line) and of the half circle of diameter $1+S_{\sigma m}$ (see text).}
\label{Principle}
\end{figure}

Let us consider now what can be deduced from an experimental photobleaching decay, knowing that the measured normalized slope, $S_{\sigma m} = \sigma^2/m +m-1$ and the intercept, $\epsilon$, contain all the available information on the initial distribution of fluorescent labels from a pbFFS experiment. Therefore, in the general case, it is impossible to independently determine the mean value and the standard deviation of the number of fluorescent labels per entity. However, the values of these statistical parameters which are compatible with a given measured slope, $S_{\sigma m}$, can be represented by $(\sigma,m)$ points located on a half circle of center $(0,\frac{1+S_{\sigma m}}{2})$ and radius $\frac{1+S_{\sigma m}}{2}$, as shown in Fig.~\ref{Principle}. This half-circle can be seen as the support of the more general $(\sigma,m)$ solution, independently of any specific kind of distribution.

The case of single-valued distributions (all entities bear the same \emph{integer} number of fluorescent labels) corresponds to discrete points along the $\sigma=0$ vertical axis. Another particular case is that of a Poisson distribution where the mean and the variance are related by $m=\sigma^2$, as depicted in Fig.~\ref{Principle}. In this case, the photobleaching decay provides all necessary information to define the distribution. The measured brightness becomes $B_{FFS}(p)=\epsilon(1+mp)$, which is the same expression as the one obtained in Ref.~\cite{Delon2010}, but it is derived here in a more general framework. Note that, for small degrees of fluorescent labelling, the percentage of unlabelled species can be very large since, according to the properties of the Poisson distribution, it is given by $e^{-m}$. 

In the general case where $S_{\sigma m} > 0$, the photobleaching slope can only provide a lower limit of the true number of entities $N$ (and hence the surface density) but no upper bound: since $N=F(1)/m\epsilon$, the condition $m \leq 1+S_{\sigma m}$ (see Fig.~\ref{Principle}) leads to a lower bound, which equals the apparent value, $N_{FFS}(1)=F(1)/B_{FFS}(1)$, corresponding to the case where all entities have the brightness $B_{FFS}(1)=(1+S_{\sigma m})\epsilon$. The fact that the real $N$ cannot be lower than the apparent $N_{FFS}$ is not new, since we argued that a distribution of brightness \emph{always} causes FFS to underestimate the number of entities (see Eq.~1 in the primary manuscript). An upper bound for $N$ can be established if we consider only \emph{fluorescent} entities: in this case, the minimum value of $m$ is larger than 1. Therefore, the true number of \emph{fluorescent} entities is included between $N_{FFS}(1)$ and $F(1)/\epsilon=(1+S_{\sigma m})N_{FFS}(1)$.

To summarize, the pbFFS method is useful if the initial fluorescent label distribution can be fully described by a limited number of degrees of freedom. When there is only one, the fluorescent label distribution can be estimated from the photobleaching slope, which makes it possible to determine the true number of entities. This is for instance the case for a Poisson law, as already used for DNA or fibrinogen fluorescent labelling \cite{Delon2010,DeMets2014}. Another example is that of a sample where all entities are uniformly fluorescently labelled. When the number of degrees of freedom of the distribution is two, it may nevertheless be possible to infer a range of values for the number of entities, by combining the constraints of the fluorescent label distribution with the relation between $\sigma$ and $m$ as established by the measured slope. This is the case encountered in the current work with SAv molecules, which are assumed to bear 1, 2 or 3 fluorescent labels. 

\section{S2. Theoretical derivation of occupancy probabilities and $m$ value ranges}

When the number of fluorescent labels borne by an entity can only take 3 non-nil values, its distribution depends on 2 independent parameters, because of the normalization. This means that for a measured parameter $S_{\sigma m}$ (that equates $\sigma^2/m +m-1$), any given possible value of $m$ fully determines the 3 occupancy probabilities. The fact that these probabilities are to be between 0 and 1 implies limited ranges of values for $m$. We now exploit this simple mathematical framework in the particular case of 1, 2 and 3 fluorescent labels, but it can be easily extended to any distribution with 3 non-nil occupancy probabilities (more generally, when the distribution of the number of fluorescent labels depends on 2 independent parameters, whatever they are, this induces constraints on $m$ that depend on the measured value of $S_{\sigma m}$).\par
Let us now consider the mean value, $m$, the standard deviation, $\sigma$ and the normalisation of the distribution. These are given by:

\begin{equation} \label{eq: m sigma p}
\begin{split}
& m = p_1 + 2p_2 + 3p_3 \\
& \sigma^2 = (1-m)^2 p_1 + (2-m)^2 p_2 + (3-m)^2 p_3 \\
& p_1 + p_2 + p_3 =1
\end{split}
\end{equation}

where $p_1$, $p_2$ and $p_3$ are the probabilities to find 1, 2 and 3 fluorescent labels. These equations can be easily transformed to express $p_1$, $p_2$ and $p_3$ as functions of $m$ and $\sigma$:

\begin{equation} \label{eq: p1p2p3}
\begin{split}
& p_1 = 3 + m(S_{\sigma m}/2 - 2)\\
& p_2 = -3 + m(3 - S_{\sigma m})\\
& p_3 = 1 + m(S_{\sigma m}/2 - 1)
\end{split}
\end{equation}

 By writing that each of these probability occupancies is between 0 and 1, one obtains the corresponding constraints on $m$ as a function of $S_{\sigma m}$:

\begin{equation} \label{eq: mmm}
\begin{split}
\frac{4}{4-S_{\sigma m}} &\leq m \leq \frac{6}{4-S_{\sigma m}}\\
 \frac{3}{3-S_{\sigma m}} &\leq m \leq \frac{4}{3-S_{\sigma m}}\\
 0 &\leq m\leq \frac{2}{2-S_{\sigma m}}
\end{split}
\end{equation}

\begin{figure}[ht!]
\centering
\includegraphics[width=0.6\textwidth]{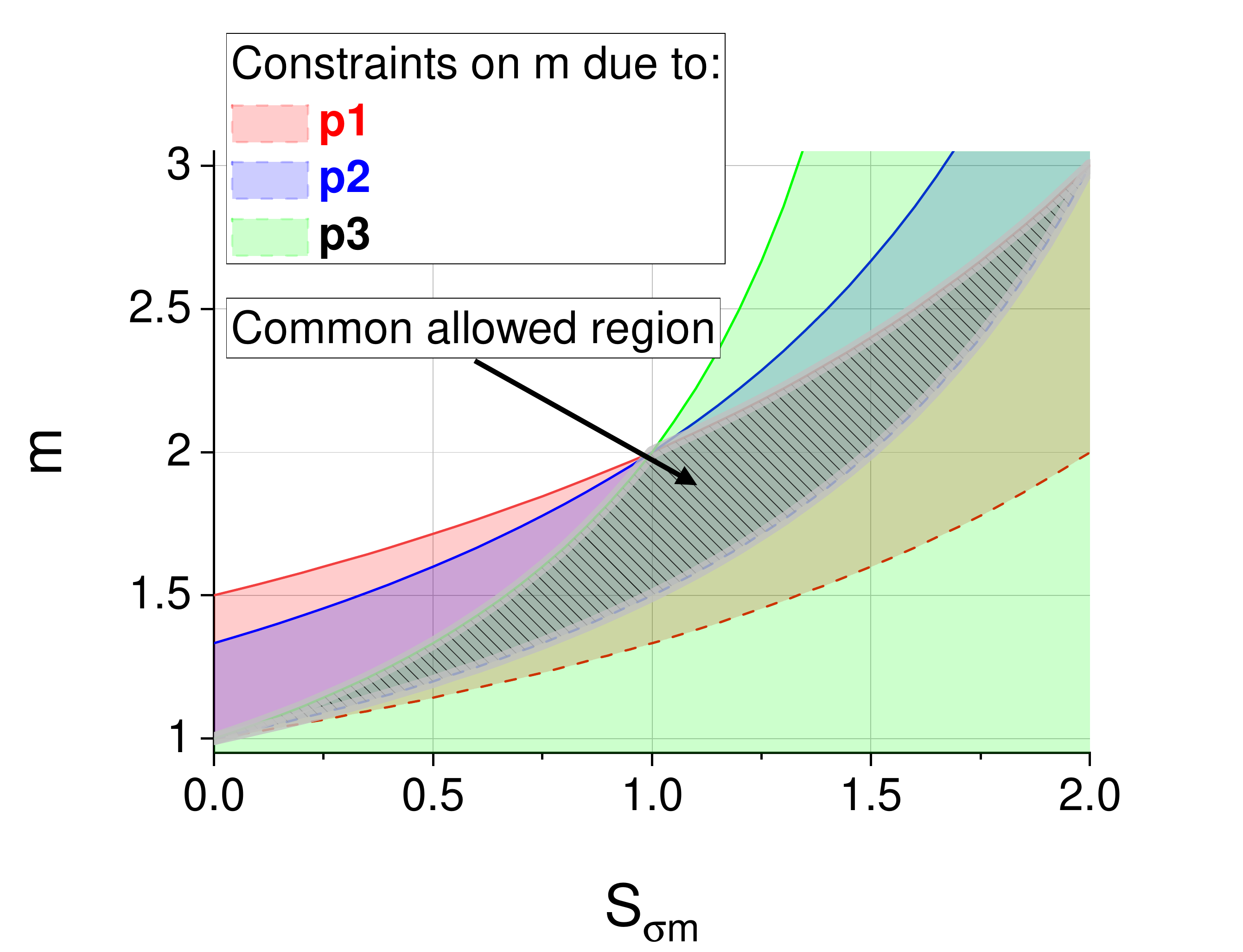}
\caption{Limits of the possible values of $m$. For each value of $S_{\sigma m}$, the occupancy probabilities for 1, 2 and 3 fluorescent labels ($p_1$ (red zone), $p_2$ (blue) and $p_3$ (green)) are to be between 0 and 1, which implies corresponding minimum and maximum values of $m$. The common zone, allowed for any of the occupancy probabilities, is hatched and surrounded in gray.}
\label{Constraints-m-S}
\end{figure}

We show in Fig.~\ref{Constraints-m-S} the series of  $m(S_{\sigma m})$ curves giving these upper and lower limits. Henceforth, the common region where all the constraints are simultaneously satisfied (hatched area) is defined by:

\begin{align}
\frac{3}{3-S_{\sigma m}} &\leq m \leq \frac{2}{2-S_{\sigma m}} \qquad \mathrm{for}\qquad 0 \leq S_{\sigma m} \leq 1\\
\frac{3}{3-S_{\sigma m}} &\leq m \leq \frac{6}{4-S_{\sigma m}} \qquad \mathrm{for} \qquad 1 \leq S_{\sigma m} \leq 2
\end{align}

It is possible to add further restrictions regarding the distribution of the number of fluorescent labels. For instance, concerning the SAv-Alex molecules the specifications of which indicate an average of 2 fluorescent labels per molecule, it is reasonable to discard the cases where $p_3 > p_2$, which leads to $m > \frac{1}{1-\frac{3}{8}S_{\sigma m}}$.

Finally, it is also interesting to ask what one would expect if one assumed that the entities could carry 2, 3 or 4 fluorescent labels. A derivation analogous to that leading to the above Eq.~\ref{eq: m sigma p} to \ref{eq: mmm} gives:

\begin{equation} \label{eq: mmmbis}
\begin{split}
& \frac{8}{5-S_{\sigma m}} \leq m \leq \frac{9}{5-S_{\sigma m}}\\
& \frac{10}{6-S_{\sigma m}} \leq m \leq \frac{12}{6-S_{\sigma m}}\\
& \frac{4}{4-S_{\sigma m}}\leq m\leq \frac{6}{4-S_{\sigma m}}
\end{split}
\end{equation}

\section{S3. Experiments with 20~nm fluorescent beads}

We performed experiments by plating 20 nm red fluorescent polystyrene beads ({FluoSpheres\texttrademark} Carboxylate Modified Microspheres, F8786, Invitrogen) on a glass substrate (Lab-Tek\texttrademark II Chambered Coverglass, Nunc) that was previously treated with $O_2$ plasma,  covered by poly(l-lysine) and then washed. After a few hours of incubation, the surface was rinsed with miliQ water to remove unbound beads. 9 different zones of 25.7$\times$25.7 $\mu m^2$ were imaged, using the same acquisition and analysis protocol as the one described in the main part of the manuscript.

\begin{figure}[ht!]
\centering
\includegraphics[width=0.9\textwidth]{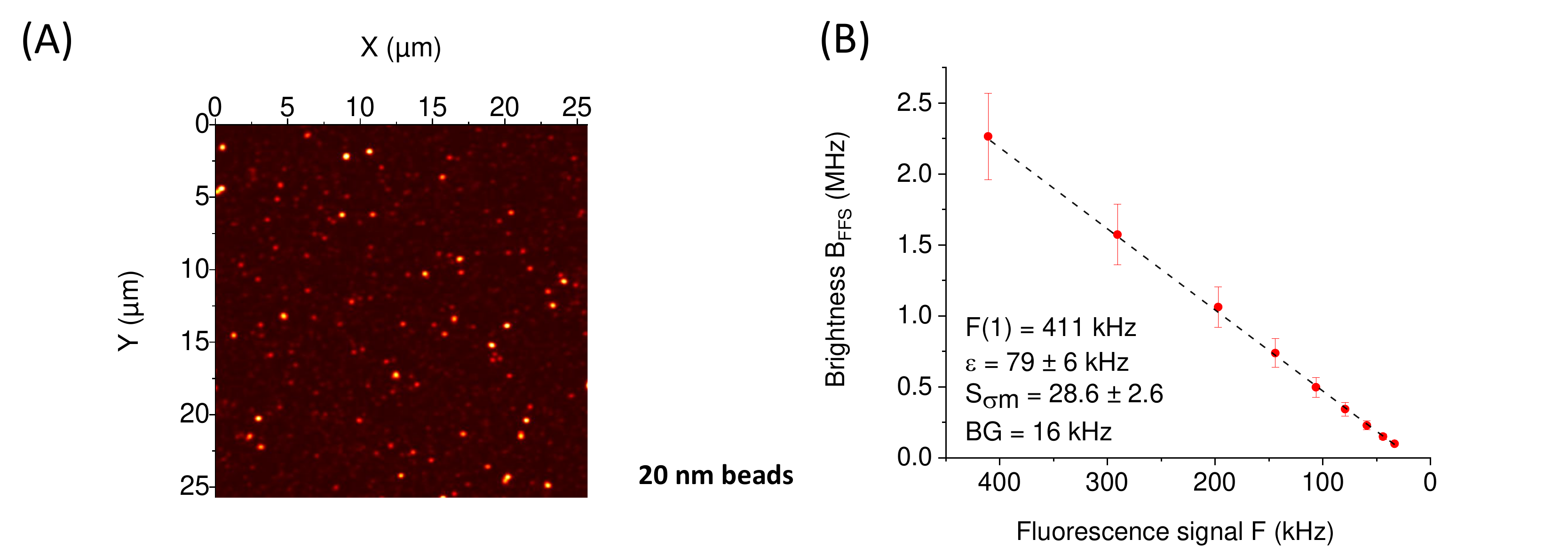}
\caption{Analysis of 20 nm fluorescent beads deposited on a glass surface. (A) Initial image (before photobleaching) of one of the 9 zones (25.7$\times$25.7 $\mu m^2$); (B) Corresponding brightness decay, the slope of which, $S_{\sigma m}$, is many tens of times larger than the one measured with SAv-Alex or bAtto molecules.}
\label{SupInfo-beads}
\end{figure}

We see in Fig.~\ref{SupInfo-beads}A an image of one zone of beads and the corresponding photobleaching decay, the different zones showing the same trend. The striking property is the typical value of the slope, $S_{\sigma m}$, of a few tens (see Fig.~\ref{SupInfo-beads}B), that is much larger than the one obtained with SAv-Alex or bAtto molecules. Exploratory single particle detection, on images acquired at sufficiently low surface concentration, has confirmed that the bead intensity distribution is very broad, as one can guess according to Fig.~\ref{SupInfo-beads}A. Assuming that the distribution of the number of fluorescent labels follows a Poisson law, we can deduce that the mean number of fluorophores per bead is also of the order of a few tens (since in this case $m$ = $S_{\sigma m}$), in agreement with the rough specifications given by the manufacturer.

We nevertheless stress the fact that the output parameters of the brightness decay fit might be very sensitive to the background. The reason relies in the very small final value of the brightness (that of the last measured point), relatively to the initial (unbleached) brightness. Depending upon the value found or fixed for the background, the estimated single fluorescent label brightness can vary a lot and so can the slope $S_{\sigma m}$.

Finally, these data show that pbFFS can in principle be applied to particles with a high degree of fluorescent labelling, although, to be reliable, measurements should be performed with a high dynamic range and a carefully controlled background.

\section{S4. Materials and methods}
\subsection{Substrate preparation}
Microscopy glass coverslips (24$\times$24 mm, Menzel Gläser) were cleaned under sonication with acetone and isopropanol and blow-dried with nitrogen. They were UV/ozone activated (UV/Ozone ProCleaner Plus, Bioforce) for 10 min, attached to a microscopy support and PLL(20)-g[3.5]-PEG(2)/PEGbiotin(3.4)50\% ($\approx$107 kDa, SuSoS AG) was incubated at 10 $\mu$g/ml in 10 mM Hepes buffer (Fisher, pH=7.2) for 45 min \cite{Huang2002}. Streptavidin ($\approx$55kDa, Sigma Aldrich), SAv, and streptavidin Alexa Fluor\texttrademark 555 conjugate ($\approx$55kDa, Molecular probes), SAv-Alex, with a labeling degree of 2 fluorophores (certificate of analysis, Molecular probes) were mixed in a ratio varying from 100:1 to 100:100 at 10 $\mu$g/ml in Hepes buffer and incubated for 30 min. A layer of biotinylated species was prepared by immobilizing Atto-labelled biotin  (Atto 565-Biotin, 921 Da, Sigma-Aldrich), bAtto, to a saturated layer of SAv, bAtto occupying the free biotin pockets on SAv. In all cases, the sample was rinsed 5 times with Hepes after incubation.

\subsection{Confocal imaging and photobleaching of surfaces}
Functionalized glass coverslips were imaged using a Leica SP8 confocal microscope with a HC PL APO 63$\times$1.2 water-immersed objective. The focal plane was determined where intensity was at maximum and then stabilized using the Adaptive Focus Control mode. The signal was detected with a hybrid detector working in the photon counting mode. An area of 25$\times$25 $\mu m^2$ with 512$\times$512 pixels was imaged 10 times with a pixel dwell time of 5 $\mu$s and a reduced laser intensity at 561 nm, so not to saturate and not to bleach the sample during image acquisitions. Then, this area was photobleached with a sufficient illumination dose (scanning time$\times$laser power) to loose roughly 30\% of the initial signal and 10 images were acquired as before. This procedure was typically repeated 6 to 8 times, in order to finish the acquisition with a remaining signal of at most 10\% of the initial one.

\subsection{Image pre-processing and ICS}
Before performing ICS analysis, it is necessary to correct the non-uniformity of the image intensity in the 1 - 10 $\mu m$ scale range because, as already discussed in \cite{DeMets2014}, it can have a strong impact on the autocorrelation function. This non-uniformity originates, either from a spatial dependence of the light efficiency of the imaging system, or from an inhomogeneous surface density. It induces long range correlations that add to the autocorrelation of interest with various detrimental effects, such as anomalous base line, width and long range behaviour. Image flattening is especially crucial when the surface density is very high and thus the autocorrelation very weak, because in this case the relative bias can be very pronounced. In order to leave the ratio of the fluctuation amplitude to the mean value unchanged, on which the estimated number of entities depends, the images are flattened by dividing them by their own smoothed version. The latter is obtained by convoluting the original image with a two-dimensional Gaussian function. The width of this Gaussian has to be small enough to damp as much as possible image inhomogeneities, but significantly larger than the radius of the ICS PSF area ($w_r \approx 0.23 \mu m$), in order not to bias the number fluctuations. Consistently with our previous study \cite{DeMets2014}, the $1/e$ half width of the Gaussian function used to smooth and flatten the images was chosen to be 2 $\mu m$. In practice, the images are individually flattened and autocorrelated. Then, for each photobleaching stage, the mean autocorrelation is fitted with Eq. S2 (see Supplementary Information), which gives a global estimation of $N_{FFS}$ and the radius $w_r$. By verifying that the latter varies by much less than 1\% during the acquisition process, we check the stability of the focus. The final goal being to analyze the variation of the brightness, $B_{FFS}$, \textit{versus} the fluorescence signal $F$ (i.e. the intensity), the images are divided in 8$\times$8 sub-images that are individually analyzed by ICS, while the value of the radius $w_r$ is set at the one estimated from the whole image, thus providing a mean and a standard deviation of the mean of the brightness at each photobleaching stage. The image processing (flattening, autocorrelation and fit) was performed using Matlab (Mathworks).

\subsection{Sample preparation and photobleaching of solutions}
To achieve photobleaching of the fluorescent labels in solution in a reasonable amount of time (a few minutes), with the available laser power ($\leq$1 mW), the solutions were confined in poly(dimethylsiloxane) (PDMS) microwells. A regular array containing pillars of 100 $\mu$m in diameter and 100 $\mu$m in height with a pitch of 400 $\mu$m was fabricated on the Si wafer. After peeling off the mold, the 2 mm thick PDMS micropatterned slabs were activated with air plasma (Atto, Diener) for 2 min to achieve a hydrophilic surface\cite{Delon2010}. Then a droplet of SAv-Alex or bAtto Hepes solution was placed on the PDMS block to enter the microwells, at initial concentrations around a few 100 nM. Such concentrations, relatively high for FCS, were necessary to saturate the microwell walls and avoid too much adsorption/desorption processes that induce unstable fluorescence signal. For control purposes, we also used solutions of Sulforhodamine B sodium salt, SRB (Sigma-Aldrich, St. Louis, USA), without further purification, diluted in either deionised water or Hepes. The PDMS block was flipped onto a Lab-Tek\texttrademark Chambered Coverglass (Nunc) to seal the microwells and thus avoiding fluid exchange with the environment. 

\subsection{FCS acquisition and fit}
The signal inside the microwells was acquired using a Nikon confocal microscope (Ti2E - A1R) with a 60$\times$water-immersed objective and a reduced laser power at 561 nm, to avoid any photobleaching during FCS acquisitions. The latter were performed using a custom made detection system, comprising a 50/50 beam splitter and a pair of avalanche photodiodes (SPCM-AQRH-44-FC, EXCELITAS) to avoid after-pulsing effects, which was connected on the auxiliary port of the microscope using a multimode optical fiber. The focal plane was set at 20 $\mu$m inside the microwells and photons were counted during 5 periods of 20~s to provide an averaged cross-correlation curve (and its corresponding standard error of the mean) using a Correlator.com software (Flex99r-12D). Proper optical adjustments and stability were controlled by fitting the diffusion time \cite{Muller2014} and measuring the brightness of a reference dye, namely sulfo-rhodamine B. Each FCS acquisition thus gives an estimation of the number of entities, $N_{FFS}$ and of the corresponding brightness, $B_{FFS}$, with a given uncertainty; it was followed by a bleaching cycle with a laser intensity adjusted to typically reduce the initial signal by 30\% before performing the next acquisition. This was repeated 5 to 8 times until about less than 10\% of the initial signal remains, in order to provide the variation of the brightness, $B_{FFS}$, with error bars, \textit{versus} the fluorescence signal $F$ (or photon count rate). 

\bibliography{ICS_photobleach}